# Multipole-mode surface solitons

Yaroslav V. Kartashov and Lluis Torner

*ICFO-Institut de Ciencies Fotoniques, and Universitat Politecnica de Catalunya,*

*Mediterranean Technology Park, 08860 Castelldefels (Barcelona), Spain*

We discover multipole-mode solitons supported by the surface between two distinct periodic lattices imprinted in Kerr-type nonlinear media. Such solitons are possible because the refractive index modulation at both sides of the interface glues together their out-of-phase individual constituents. Remarkably, we find that the new type of solitons may feature highly asymmetric shapes and yet they are stable over wide domains of their existence, a rare property to be attributed to their surface nature.

Nonlinear periodic structures, or optical lattices, may support stable solitons that have no analogue, or that are highly unstable, in uniform nonlinear media. Salient examples include scalar gap, vortex, and multipole-mode solitons. Recent achievements in optical lattice induction [1-4] enabled experimental observation of such solitons. Thus, the properties of vortex lattice solitons were analyzed in Ref. [5] and their experimental observation was reported in [6,7]. Another interesting class of solutions, i.e. multipole-mode solitons, was studied theoretically in Refs. [8,9] and observed experimentally in [10,11]. Bessel optical lattices also support stable multipole-mode solitons [12-14]. It was recently predicted and verified experimentally that nonlinear interfaces between periodic structures and uniform media also support solitons [15-17]. Such *surface solitons* are located at the very interface and their properties depend crucially on the ratio of refractive index of the uniform medium and mean refractive index of the periodic structure. Suitable periodic structures with engineered properties can be made in waveguide arrays with technologies currently available, and optical induction of the lattices is a potential future alternative. The power level required for existence of surface solitons at such interfaces may be substantially reduced in comparison with that for interfaces of uniform media (see Refs. [18,19] for reviews) because of tunable, shallow refractive index modulations which are possible in optically-induced lattices or waveguide arrays. Recently, gap surface solitons were predicted [20] and experimentally observed [21]. New types of interfaces can be produced by different lattices, opening new possibilities for exploration. In this Letter we consider the interface between two periodic lattices and show that they support multipole-mode solitons that have no analogues in uniform materials. In contrast to multipole-mode solitons in infinite periodic lattices, surface multipole solitons feature strongly asymmetric shapes. Yet, we found them to be stable in wide domains of their existence. Our findings constitute the first known example of a nontrivial soliton structure guided by a surface.

We thus address beam propagation at the interface produced by two lattices imprinted in a focusing medium with Kerr-type saturable nonlinearity, governed by the nonlinear Schrödinger equation for dimensionless complex amplitude of light field $q$:

$$i\frac{\partial q}{\partial \xi} = -\frac{1}{2}\left(\frac{\partial^2 q}{\partial \eta^2} + \frac{\partial^2 q}{\partial \zeta^2}\right) - \frac{q|q|^2}{1+S|q|^2} - R(\eta,\zeta)q. \qquad (1)$$

Here the transverse $\eta$ and longitudinal $\xi$ coordinates are scaled to the beam width and diffraction length, respectively, and $S$ is the saturation parameter. The function $R(\eta,\zeta) = \delta p H(\eta) + (p/4)[1-\cos(\Omega\eta)][1-\cos(\Omega\zeta)]$ describes the refractive index profile, where $p$ is the depth of periodic part of the lattice, $\Omega$ is its frequency, the function $H(\eta) \equiv 0$ for $\eta \leq 0$, and $H(\eta) \equiv 1$ for $\eta > 0$, and $\delta p$ characterizes the height of step in the constant part of refractive index. Such refractive index landscapes might be optically-induced in photorefractive crystals, where the lattice might be created by interfering four plane waves, while a non-uniform incoherent background illumination can produce a step in the refractive index profile at $\eta = 0$. Other types of lattice interfaces could be realized by applying different voltages across the crystal. Equation (1) holds provided that intensities of soliton and lattice-creating beams are small compared with the background illumination level and crystal is biased with a strong static electric field. We stress the experimental feasibility of such setting, as well as the possibility of tuning the properties of the interface. The model (1) holds also for nonlinear media with an imprinted refractive index modulation (see Ref. [22,23] for observation of discrete solitons in waveguide arrays imprinted in saturable $LiNbO_3$ crystal and AlGaAs arrays). Among the conserved quantities of Eq. (1) is the energy flow $U = \int \int_{-\infty}^{\infty} |q|^2 \, d\eta d\zeta$.

We search for solitons with the form $q = w(\eta,\zeta)\exp(ib\xi)$, where $w$ describes the field distribution and $b$ is the propagation constant. The soliton profiles were found numerically with a standard relaxation method. To elucidate their linear stability we searched for perturbed solutions $q = (w + u + iv)\exp(ib\xi)$ of Eq. (1), where $u(\eta,\zeta,\xi)$ and $v(\eta,\zeta,\xi)$ are real and imaginary parts of perturbation that can grow with complex rate $\delta$ upon propagation. Linearization of Eq. (1) around $w$ yields the system of equations

$$\frac{\partial v}{\partial \xi} = \frac{1}{2}\left(\frac{\partial^2 u}{\partial \eta^2} + \frac{\partial^2 u}{\partial \zeta^2}\right) - bu + \frac{3w^2 + Sw^4}{(1+Sw^2)^2}u + pRu,$$
$$\frac{\partial u}{\partial \xi} = -\frac{1}{2}\left(\frac{\partial^2 v}{\partial \eta^2} + \frac{\partial^2 v}{\partial \zeta^2}\right) + bv - \frac{w^2 + Sw^4}{(1+Sw^2)^2}v - pRv, \qquad (2)$$

that we solved numerically. We set $\Omega = 4$, $S = 0.05$, and vary parameters $b$, $p$, and $\delta p$.

The simplest asymmetric multipole-mode solitons supported by the interface consist of two bright spots with $\pi$ phase-jump between them that are located at different sides of the interface (Fig. 1). Notice that such *solitons cannot exist* at the interface between uniform materials because of repulsive forces acting between out-of-phase spots. Periodic refractive index modulation can compensate repulsive forces acting between out-of-phase soliton constituents even if they have different peak amplitudes and results in multipole soliton formation. Importantly, despite the fact that topological (phase) structure of multipole surface solitons is similar to that for multipole solitons in periodic lattices [8,9] and twisted unstaggered modes in waveguide arrays introduced in [24] and further classified in [25], their intensity distribution can be strongly asymmetric, since lattices forming the interface require different peak powers for self-sustained light beam propagation for a given propagation constant $b$. The energy flow of dipole-mode surface soliton is a non-monotonic function of propagation constant (Fig. 2(a)). We found that $U$ diverges as $b$ approaches cutoff $b_{co}$, and that far from the cutoff $U$ grows with $b$. In clear contrast to multipole solitons in usual periodic lattices [8,9], right spot of surface soliton strongly expands over the lattice at $\eta > 0$ and $\max q|_{\eta>0} \to 0$, while its left spot remains well localized as $b \to b_{\text{co}}$ (Fig. 1(a)). The asymmetry in soliton profile becomes less pronounced with increase of $b$ (Fig. 2(a)).

A similar behavior was encountered for other types of multipole surface solitons, including quadrupole ones, that consist of two bright spots in the region $\eta \leq 0$ and two spots at $\eta > 0$, while soliton phase changes by $\pi$ between different quadrants of $(\eta, \zeta)$ plane (Figs. 1(c) and 1(d)). The power distribution $W = U_{\text{r}}/U_{\text{l}}$ defined as the ratio between the power $U_{\text{r}}$ localized at $\eta > 0$ and the power $U_{\text{l}}$ localized at $\eta \leq 0$, diverges as $b \to b_{\text{co}}$, since $U_{\text{r}}(b \to b_{\text{co}}) \to \infty$, while $U_{\text{l}}$ remains finite (Fig. 2(b)). The quantity $W$ characterizes the degree of asymmetry of light intensity distribution of surface soliton. Notice the clearly pronounced minimum in the dependence $W(b)$. Figure 1(c) shows that $W_{\min}$ decreases with increase of the height $\delta p$ of step in refractive index, which indicates progressive asymmetry growth.

We found that for fixed lattice depth cutoff $b_{\text{co}}$ increases almost linearly with $\delta p$ (Fig. 2(d)). In contrast, for fixed $\delta p$ the dependence $b_{\text{co}}(p)$ is nonmonotonic, as $b_{\text{co}}$ tend to infinity for $p \to 0$ and $p \to \infty$. Figure 2(e) shows part of the dependence $b_{\text{co}}(p)$ for $p > 3$. Importantly, linear stability analysis revealed that surface solitons become completely stable when propagation constant exceeds a critical value $b_{\text{cr}} > b_{\text{co}}$. The width of instability domain remains almost unchanged with increase of $\delta p$, i.e. with increase of asymmetry in soliton profile (Fig. 2(d)), and it decreases with increase of lattice depth $p$ (Fig. 2(e)). Thus even *moderate periodic refractive index modulation can stabilize highly asymmetric multipole surface solitons*. We found that similarly to weakly-localized twisted modes in waveguide arrays [24], the largest part of instability domain for surface multipole solitons is associated with oscillatory instabilities ($\text{Re}(\delta)\text{Im}(\delta) \neq 0$), while close to cutoff surface solitons are exponentially unstable (Fig. 2(f)). Notice that oscillatory instability in the low-power limit is a characteristic signature of multipole solitons existing in different physical systems. Increasing the lattice depth suppresses the instability and it is accompanied by reduction of maximal value of $\text{Re}(\delta)$.

The results of linear stability analysis were fully confirmed by direct simulations of Eq. (1). We used input conditions with $q|_{\xi=0} = w(\eta, \zeta)[1 + \rho(\eta, \zeta)]$, where $\rho(\eta, \zeta)$ describes white noise with Gaussian distribution and variance $\sigma_{\text{noise}}^2$. In all cases solitons with $b > b_{\text{cr}}$ perturbed with considerable input noise survived over huge distances (Fig. 3). Decay of unstable solitons typically results in emission of part of power localized in the region $\eta > 0$ and formation of soliton in the region $\eta \leq 0$. Simulations show that asymmetric surface solitons can be excited by Gaussian beams with properly selected amplitudes and phases, which confirms the robustness of the solitons. Finally, besides dipole and quadrupole solitons, we have also found a variety of more complex stable asymmetric solitons (not shown here) supported by the interface between two lattices.

Summarizing, we have reported on the first known example of stable, nontrivial soliton structures guided an interface between two optical lattices. We found that such solitons can be stable and robust in lattices with moderate depths even if they feature highly asymmetric profiles. The setting explored can be implemented experimentally with optical-induction and in suitable waveguide arrays, and hence it opens a route to the observation of rich variety of nonlinear surface waves, at reduced power levels.

# References with titles

12. Y. V. Kartashov, V. A. Vysloukh, and L. Torner, "Rotary solitons in Bessel optical lattices" Phys. Rev. Lett. **93**, 093904 (2004).
13. Y. V. Kartashov, A. A. Egorov, V. A. Vysloukh, and L. Torner, "Rotary dipole-mode solitons in Bessel optical lattices" J. Opt. B **6**, 444 (2004).
14. Y. V. Kartashov, A. A. Egorov, V. A. Vysloukh, and L. Torner, "Stable soliton complexes and azimuthal switching in modulated Bessel optical lattices" Phys. Rev. E **70**, 065602(R) (2004).
15. K. G. Makris, S. Suntsov, D. N. Christodoulides, G. I. Stegeman, and A. Hache, "Discrete surface solitons," Opt. Lett. **30**, 2466 (2005).
16. J. Hudock, S. Suntsov, D. N. Christodoulides, and G. I. Stegeman, "Vector discrete nonlinear surface waves," Opt. Express **13**, 7720 (2005).
17. S. Suntsov, K. G. Makris, D. N. Christodoulides, G. I. Stegeman, A. Hache, R. Morandotti, H. Yang, G. Salamo, and M. Sorel, "Observation of discrete surface solitons," Phys. Rev. Lett. **96**, 063901 (2006).
18. "Nonlinear surface electromagnetic phenomena," Ed. by H. E. Ponath and G. I. Stegeman, North Holland, Amsterdam (1991).
19. G. I. Stegeman, E. M. Wrigth, N. Finlayson, R. Zanoni, and C. T. Seaton, "Third order nonlinear integrated optics," J. Lightwave Technol. **6**, 953 (1988).
20. Y. V. Kartashov, V. A. Vysloukh, and L. Torner, "Surface gap solitons," Phys. Rev. Lett. **96**, 073901 (2006).
21. C. R. Rosberg, D. N. Neshev, W. Krolikowski, A. Mitchell, R. A. Vicencio, M. I. Molina, and Y. S. Kivshar, "Observation of surface gap solitons in semi-infinite waveguide arrays," arXiv: physics/0603202 (2006).
22. F. Chen, M. Stepic, C. E. Rüter, D. Runde, D. Kip, V. Shandarov, O. Manela, and M. Segev, "Discrete diffraction and spatial gap solitons in photovoltaic $LiNbO_3$ waveguide arrays," Opt. Express **13**, 4314 (2005).
23. J. Meier, G. Stegeman, D. Christodoulides, R. Morandotti, M. Sorel, H. Yang, G. Salamo, J. Aitchison, and Y. Silberberg, "Nonlinear beam interactions in 1D discrete Kerr systems," Opt. Express **13**, 1797 (2005).
24. S. Darmanyan, A. Kobyakov, and F. Lederer, "Stability of strongly localized excitations in discrete media with cubic nonlinearity," JETP **86**, 682 (1998).
7

# References without titles

# Figure captions

Figure 1. Field modulus distributions of dipole-mode surface solitons corresponding to $b = 3.45$ (a) and $b = 4.05$ (b) at $p = 4$, $\delta p = 2$ and quadrupole-mode solitons corresponding to $b = 5.33$ (c) and $b = 7.5$ (d) at $p = 4$, $\delta p = 4$. Vertical dashed lines indicate interface position.

Figure 2. Total energy flow (a) and power distribution (b) vs propagation constant for dipole-mode soliton at $p = 4$ and $\delta p = 2$. Points marked by circles in (a) correspond to solitons shown in Figs. 1(a) and 1(b). (c) Energy sharing vs $\delta p$ at $p = 4$. (d) Domains of stability and instability (shaded) on $(\delta p, b)$ plane at $p = 6$ (d) and on $(p, b)$ plane at $\delta p = 4$ (e). (f) Real part of perturbation growth rate vs propagation constant at $p = 4$ and $\delta p = 4$.

Figure 3. Stable propagation of dipole-mode soliton with $b = 7$ and quadrupole-mode soliton with $b = 7.5$ in the presence of white input noise with variance $\sigma_{\text{noise}}^2 = 0.01$. The field modulus distributions are shown at $\xi = 0$ (a),(c) and $\xi = 256$ (b),(d). In all cases $p = 4$, $\delta p = 4$.

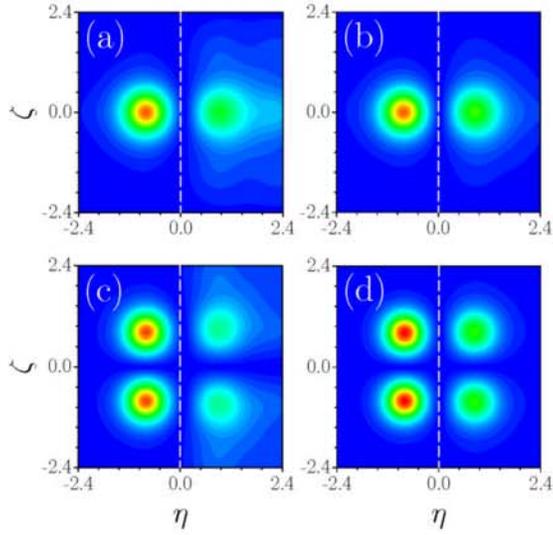

Figure 1. Field modulus distributions for dipole-mode surface solitons corresponding to $b = 3.45$ (a) and $b = 4.05$ (b) at $p = 4$, $\delta p = 2$ and quadrupole-mode solitons corresponding to $b = 5.33$ (c) and $b = 7.5$ (d) at $p = 4$, $\delta p = 4$. Vertical dashed lines indicate interface position.

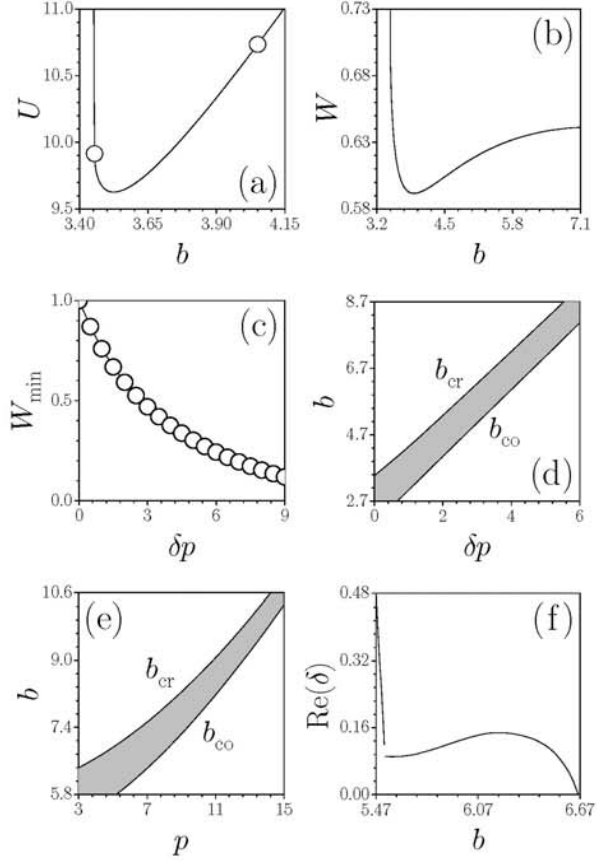

Figure 2. Total energy flow (a) and power distribution (b) vs propagation constant for dipole-mode soliton at $p=4$ and $\delta p = 2$. Points marked by circles in (a) correspond to solitons shown in Figs. 1(a) and 1(b). (c) Energy sharing vs $\delta p$ at $p=4$. (d) Domains of stability and instability (shaded) on $(\delta p, b)$ plane at $p=6$ (d) and on $(p,b)$ plane at $\delta p = 4$ (e). (f) Real part of perturbation growth rate vs propagation constant at $p=4$ and $\delta p = 4$.

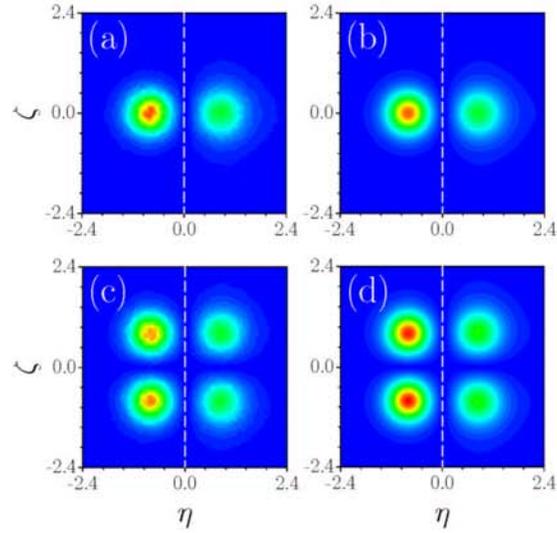

Figure 3. Stable propagation of dipole-mode soliton with $b = 7$ and quadrupole-mode soliton with $b = 7.5$ in the presence of white input noise with variance $\sigma^2_{\text{noise}} = 0.01$. The field modulus distributions are shown at $\xi = 0$ (a),(c) and $\xi = 256$ (b),(d). In all cases $p = 4$, $\delta p = 4$.